\documentclass[reprint,amsmath,amssymb,aps,superscriptaddress]{revtex4-2}
\usepackage{graphicx}
\usepackage{dcolumn}
\usepackage{bm}

\usepackage[normalem]{ulem}
\usepackage[colorlinks,citecolor=blue,urlcolor=blue,bookmarks=false,hypertexnames=true]{hyperref} 
\usepackage{xcolor}
\definecolor{codegreen}{rgb}{0,0.6,0}

\newcommand{\comt}[1]{}

\newcommand{\apx}[1]{Appendix~\ref{#1}}
\newcommand{\eq}[1]{Eq.~(\ref{#1})}
\newcommand{\fig}[1]{Fig.~\ref{#1}}

\newcommand{\ket}[1]{|#1 \rangle}
\newcommand{\bra}[1]{\langle #1|}

\newcommand{\tr}{\text{Tr}}
\newcommand{\J}{\mathcal{J}}
\newcommand{\M}{\mathcal{M}}
\newcommand{\id}{\openone}

\newcommand{\W}{\mathcal{W}}

\newcommand{\blue}[1]{{\color{blue} #1}}

\newcommand{\ketbra}[2]{\left|#1\right\rangle\left\langle#2\right|}

\begin{document}

\preprint{APS/123-QED}

\title{Revealing non-Markovian Kondo transport with waiting time distributions}

\author{Feng-Jui Chan}
\email{These authors contributed equally to this work.}
\affiliation{Department of Physics, National Cheng Kung University, 701 Tainan, Taiwan}
\affiliation{Center for Quantum Frontiers of Research and Technology, NCKU, 701 Tainan, Taiwan} 
\author{Po-Chen Kuo}
\email{These authors contributed equally to this work.}
\affiliation{Department of Physics, National Cheng Kung University, 701 Tainan, Taiwan}
\affiliation{Center for Quantum Frontiers of Research and Technology, NCKU, 701 Tainan, Taiwan}
\author{Neill Lambert}
\email{nwlambert@gmail.com}
\affiliation{Theoretical Quantum Physics Laboratory, Cluster for Pioneering Research, RIKEN, Wakoshi, Saitama 351-0198, Japan}
\author{Mauro Cirio}
\email{cirio.mauro@gmail.com}
\affiliation{Graduate School of China Academy of Engineering Physics, Haidian District, Beijing, 100193, China}
\author{Yueh-Nan Chen}
\email{yuehnan@mail.ncku.edu.tw}
\affiliation{Department of Physics, National Cheng Kung University, 701 Tainan, Taiwan}
\affiliation{Center for Quantum Frontiers of Research and Technology, NCKU, 701 Tainan, Taiwan}
\affiliation{Physics Division, National Center for Theoratical Sciences, Taipei, 106319, Taiwan}

\date{\today}

\begin{abstract}
We investigate non-Markovian transport dynamics and signatures of the Kondo effect in a single impurity Anderson model. The model consists of a quantum dot (QD) with ultra-strong coupling to a left lead and weak coupling to a right lead acting as a detector. We calculate the waiting time distribution (WTD) of electrons tunneling into the detector using a combination of the hierarchical equations of motion approach (HEOM) and a dressed master equation. Oscillations emerge in the short-time WTD, becoming more pronounced with stronger left-lead coupling. Fourier analysis reveals a blue shift in the oscillation frequency as coupling increases, indicating enhanced system-bath hybridization.  Crucially, comparison with a dressed master equation confirms that these oscillations are a direct consequence of non-Markovian system-bath correlations.  We examine the Kondo effect's influence on these oscillations by varying the quantum dot's Coulomb repulsion. Increasing this interaction enhances the WTD oscillations, coinciding with the signatures of a strengthened Kondo resonance in the quantum dot's density of states. Our results demonstrate that WTD oscillations offer a valuable tool for probing non-Markovian system-bath interactions and the emergence of Kondo correlations within quantum dot systems.

\end{abstract}

\maketitle 

\section{Introduction}

Open quantum systems, where a system of interest is coupled to one or more environments, exhibit rich dynamics that often deviate from the traditional Markovian picture~\cite{Breuer2002,Daniel2020}. In the Markovian approximation, the system's evolution is described by a memory-less master equation, such as the Lindblad equation~\cite{Davies1976, Carmichael1991, Breuer2007}. However, this approximation breaks down when the system-environment coupling is strong or when the environmental correlation times become long, leading to non-Markovian dynamics~\cite{Liu2011, Zhang2012, Xiong2015, Breuer2016, Vega2017}.

Non-Markovian dynamics are characterized by the backflow of information from the environment to the system, giving rise to memory effects~\cite{Wolf2008, Breuer2009, Laine2010, Liu2019}. These effects play a crucial role in various quantum processes, including quantum transport in nanoscale devices~\cite{Braggio2006, Flindt2008, Schaller2009, Moreira2020}. Understanding and controlling non-Markovian effects is essential for harnessing the full potential of quantum technologies, as they can profoundly impact the coherence and entanglement properties of quantum systems~\cite{Bellomo2007, Rivas2010, Cialdi2011, Huelga2012, Xu2013, Orieux2015}.

One striking manifestation of strong system-environment coupling is the Kondo effect~\cite{Kouwenhoven2001}, which arises from the many-body entanglement between electrons in the system and those in the environment~\cite{Cho2006, Amico2008, Affleck2009, Lee2015, Wagner2018, Kim2021, Kuo2023}. In particular, QDs have emerged as versatile platforms for engineering and probing the Kondo effect~\cite{Sprinzak2002, Kalish2004, Pustilnik_2004, Keller2014, LeHur2015, Shang2018, Kuo2023}, offering insights into the interplay between many-body physics and quantum transport~\cite{Jeong2001, Komnik2005, Gogolin2006, Grobis2008, Nguyen2020}.

A powerful tool for studying the influence of non-Markovian environments and the Kondo effect on quantum transport~\cite{Grobis2008, Zheng2009, Schiro2019, Tang2014_2, Nguyen2020, Kuo2023} is the WTD~\cite{Brandes2008, Welack2008, Welack2009, Welack2009Jchem, Thomas2013, Tang2014_1, Sothmann2014, Talbo2015, Rudge2016_1, Rudge2016_2, Ptaszynski2017, Rudge2018, Stegmann2018, Kleinherbers2018, Tang2018, Rudge2019, Kleinherbers2021, Stegmann2021}, which quantifies the probability of observing electron transfer events at different time intervals. the WTDs have been extensively used to analyze electron transport in QDs~\cite{Brandes2008, Welack2008, Welack2009, Welack2009Jchem, Thomas2013, Kleinherbers2021, Fu2022, Kleinherbers2023}, revealing coherent oscillations~\cite{Brandes2008, Kleinherbers2021}, entanglement signatures~\cite{Kleinherbers2021}, identifying the Majorana states for the Majorana island device~\cite{Fu2022}, and the influence of non-Markovian environments~\cite{Thomas2013}.

Despite advances in understanding non-Markovian transport, its interplay with the Kondo effect in shaping the WTD remains largely unexplored. This work aims to bridge this gap by studying a single impurity Anderson model (SIAM), describing a QD ultra-strongly coupled to a (left) non-Markovian lead and weakly coupled to a detector (right) Markovian lead. This allows us to combine the HEOM, describing the interaction with the left lead, with a master equation to describe the right detector lead and allow us to obtain the WTD.  

This approach reveals the emergence of short-time oscillations in the WTD. These oscillations become more pronounced when increasing the left-lead coupling.  Fourier analysis reveals this stronger coupling induces a blue shift in the frequency domain, signifying the enhanced system-environment interactions.  Crucially, these oscillations are absent when using the dressed master equation alone, confirming their non-Markovian origin.

Furthermore, we explore the Kondo resonance's influence on the WTD.  Increasing Coulomb interactions within the dot (impurity) strengthens the Kondo resonance in the density of states (DOS), as expected.  Correspondingly, the WTD oscillations are amplified, demonstrating an indirect link between the Kondo effect and non-Markovian transport dynamics.

\begin{figure}[!htbp]
    \centering
    \includegraphics[width=1\linewidth]{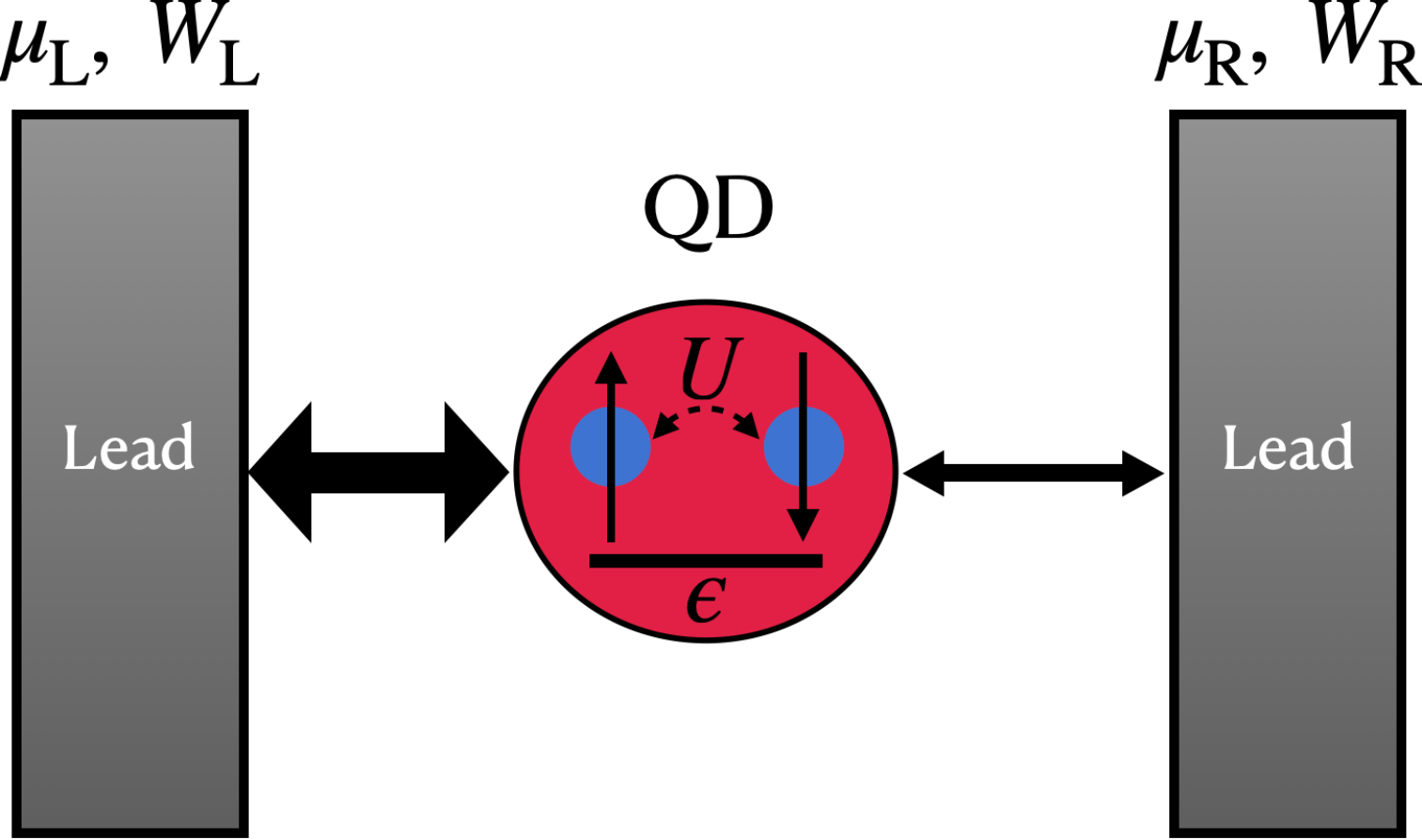}
    \caption{Schematic representation of the SIAM considered in this study. The model depicts a single quantum dot (QD) with spin configurations, ultrastrongly coupled to a left bath (L) and weakly coupled to a right bath (R). The energy of the electronic state in the QD is denoted by $\epsilon$. $U$ represents the Coulomb repulsion energy between two electrons with opposite spins occupying the QD. Here, $W_{\alpha}$ and $\mu_{\alpha}$ represent the bandwidth and chemical potential of the lead ($\alpha = \text{L, R}$), respectively.
    }
    \label{fig:SIAM}
\end{figure}

\section{Model}

We consider a SIAM as depicted in \fig{fig:SIAM}. The system is comprised of a QD (the single impurity) coupled to two leads. The QD can be occupied by either a single electron or two electrons with different spin configurations $\sigma= \uparrow,\downarrow$ (up, down). The total  Hamiltonian of the system and leads is ($\hbar = 1$)
\begin{equation}\label{eq:total_Ham}
\begin{aligned}
    H_\mathrm{T} = H_\mathrm{s} + H_\mathrm{f} + H_\mathrm{sf}.
\end{aligned}
\end{equation}
Here, the impurity is described by the system Hamiltonian
\begin{equation}
    H_\mathrm{s} = \sum_{\alpha=\uparrow, \downarrow}
    \epsilon \hat{n}_\sigma + U \hat{n}_\uparrow \hat{n}_\downarrow,
\end{equation}
where the operator $\hat{n}_\sigma = d_\sigma^\dagger d_\sigma$ describes the number of electrons with spin $\sigma$ in the QD, each associated with the annihilation operator $d_\sigma$ and the energy-level scale $\epsilon$. The Coulomb interaction between two electrons has a strength given by the repulsion energy $U$ and is described by the non-linear operator $\hat{n}_\uparrow \hat{n}_\downarrow$.

The leads are described by the Hamiltonian
\begin{equation}
    H_\mathrm{f}=\sum_{k,\alpha,\sigma}\omega_{k,\alpha,\sigma}
    c_{k,\alpha,\sigma}^\dagger c_{k,\alpha,\sigma},
\end{equation}
where
$c_{k,\alpha,\sigma}$ creates an electron in the state $k$ of the $\alpha$-th lead ($\alpha \in \{\mathrm{L}, \mathrm{R} \}$). The tunneling between the QD and the leads is given by
\begin{equation}
    H_\mathrm{sf} = \sum_{k,\alpha,\sigma} g_{k,\alpha,\sigma} \left( c_{k,\alpha,\sigma}^\dagger d_\sigma + c_{k,\alpha,\sigma} d_\sigma^\dagger \right),
\end{equation}
    parametrized by the coupling strengths $g_{k,\alpha,\sigma}$. In the continuum limit, these coefficients can be encoded in a spectral density function for which we choose a Lorentzian lineshape

\begin{equation}
    J_\mathrm{\alpha}(\omega) = \frac{1}{2\pi}
    \frac{\Gamma_\alpha W_\alpha^2}{\left( \omega - \mu_\alpha \right)^2 + W_\alpha^2}. 
\end{equation}
Here, $\Gamma_\alpha$ describes the overall coupling strength between the QD and the $\alpha$-lead, while $W_\alpha$ and $\mu_{\alpha}$ represent the bandwidth and the chemical potential of the $\alpha$-lead, respectively.

In this work, we assume the system to be ultra-strongly coupled to the left lead and only weakly coupled to the right one, which can thereby be used as a detector with a well-defined WTD.
The effects of the fermionic leads on the system are fully encoded in the two-time correlation functions

\begin{equation}
\label{eq:corr_main}
\begin{aligned}
    C_{\alpha}^\nu(t) & = \tr_\mathrm{f} \left[ \sum_k \Gamma_{\alpha, k}^2  d_\sigma^{\nu} d_\sigma^{\bar{\nu}} \rho_\mathrm{f}(0) \right] e^{\nu i \omega t}\\
    & =\frac{1}{2\pi}\int_{-\infty}^\infty d\omega J_\alpha(\omega) \left[ \frac{1-\nu}{2} + \nu n_\alpha^\mathrm{eq}(\omega) \right]e^{\nu i \omega  t},
\end{aligned}
\end{equation}
in terms of the Fermi-Dirac distribution $n_\alpha^\mathrm{eq}(\omega)=\{\mathrm{exp}[(\omega-\mu_\alpha)/k_\mathrm{B}T] +1 \}^{-1}$ at temperature $T$, where $k_\mathrm{B}$ is the Boltzmann constant. Here $\nu = \pm1$ characterizes the distinction between particles and holes, and $d^{\nu=1}_\sigma$ = $d^\dagger_\sigma$ and $d^{\nu=-1}_\sigma = d_\sigma$ (with $\bar{\nu}=-\nu$). 
While the correlations in \eq{eq:corr_main} cannot in general be written in a closed analytical form, it is usually advantageous to consider the following ansatz

\begin{equation}
    C_{\alpha}^\nu(t) = \sum_{l=0}^{l_\mathrm{max}}\eta_{l}^\nu \mathrm{exp}[-\gamma_{\alpha,\nu,l} (t)].
\end{equation}
This representation expresses the correlation as an exponential series whose strength-coefficients  $\eta_{l}^\nu$ and decay rates $\gamma_{\alpha,\nu,l}$ can be found using the Pad\'e spectral decompositions as in \cite{Jie2011,Huang2023}.
\begin{figure*}[!htbp]
    \centering
    \includegraphics[width=1\linewidth]{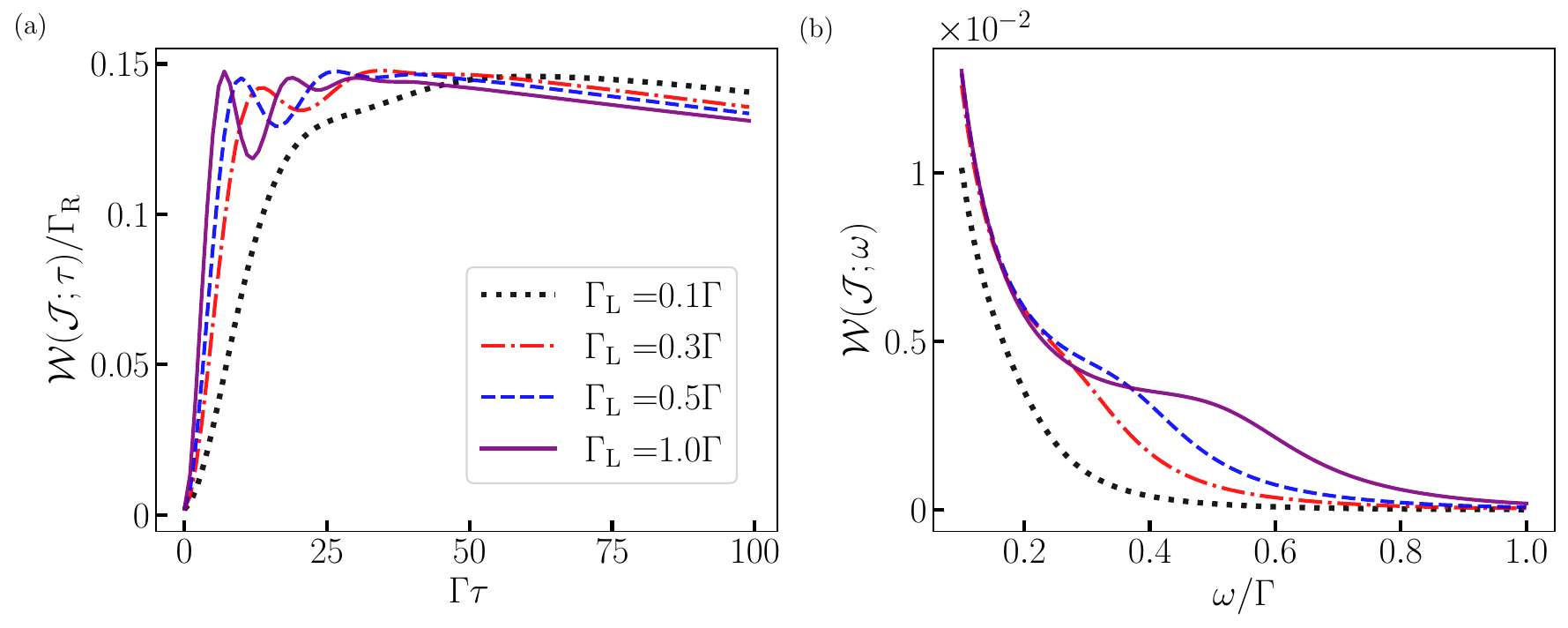}
    \caption{
    (a) $W(\J;\tau)/\Gamma_{\text{R}}$ as a function of time $\Gamma t$ for different values of $\Gamma_\mathrm{L}$ with $U = 1\Gamma$, where $\Gamma_\mathrm{R}=0.01\Gamma$. The WTD exhibits oscillations that become more pronounced as $\Gamma_\mathrm{L}$ increases. (b) Spectrum of the WTD, $\W(\J;\omega)$, obtained via the Fourier transform of $\W(\J;\tau)$ (\eq{eq:WTD_fourier}). As $\Gamma_\mathrm{L}$ increases, the WTD spectrum shows not only an increase in the oscillation amplitude but also a blue shift in the oscillation frequency.
    }
    \label{fig:WTD_coupling_spin}
\end{figure*}

In order to capture the non-perturbative effects originating from the strong interaction with the left lead, we employ the HEOM approach \cite{Yan2012,lambert2020bofinheom,Huang2023} which is, in principle, numerically exact  \cite{Numericallyexact2020} as it goes beyond the standard master equations based on the Born-Markov approximation.

This enables us to comprehensively explore the interplay between system-bath correlations and non-Markovian transport characteristics. 
Explicitly, the HEOM for the strongly coupled QD-(left lead) system can be written as
\begin{equation}
\label{eq:HEOM_1}
   \partial_t \rho_\mathrm{\textbf{j}}^{(m,p)}(t) 
   \equiv \hat{\M}\rho_\textbf{j}^{(m,p)}(t)\;,
\end{equation}
in terms of the HEOM Liouvillian superoperator $\hat{\mathcal{M}}$, which characterizes the dynamics of the set of auxiliary density operators (ADOs) defined by $\rho_\mathrm{\textbf{j}}^{(m,p)}(t)$. Here, the pair $(m,p)$ represents the $m$th level fermionic ADO with parity $p$, and \textbf{j} denotes the vector $[j_m,\cdots,j_1]$, where each $j$ represents a specific ensemble with multi-index $\{ \alpha, v, h, \sigma_\mathrm{f}\}$ in which $\sigma_\mathrm{f}$ specifies the remaining system quantum numbers. The superoperator in \eq{eq:HEOM_1} can be explicitly defined by its action on the auxiliary density matrices as
\begin{equation}
\begin{array}{lll}
\label{eq:HEOM_2}
\hat{\M}\rho_\textbf{j}^{(m,p)}(t) &=&\displaystyle -\left( i \hat{\mathcal{L}}_\mathrm{s} + \sum_{w=1}^m \gamma_{q_w} \right)\rho_{j}^{(m,p)}(t)\\
&&\displaystyle-i\sum_{j' \notin \textbf{j}} \hat{\mathcal{A}}_{j'}\rho_{\textbf{j}^+}^{(m+1,p)}(t)\\ 
   &&\displaystyle-i \sum_{w=1}^m (-1)^{m-w}\hat{B}_{q_w}\rho_{\textbf{j}_w^-}^{(m-1,p)}(t).
\end{array}
\end{equation}
Here, $\hat{\mathcal{L}}_\mathrm{s}[\cdot]=[H_\mathrm{s}(t),\cdot]$ describes the system dynamics while $\hat{\mathcal{A}}_j$ and $\hat{\mathcal{B}}_j$ are the fermionic superoperators describing the system-bath interaction. Specifically, they
couple the $m$th-level-fermionic ADOs to the $(m+1)$th-level- and $(m-1)$th-level-fermionic ADOs and they are explicitly given by 
\begin{equation}
\begin{aligned}
    \hat{\mathcal{A}}_j[\cdot] &=(-1)^{\delta_{p,-}}
    \lbrace d_{\sigma_\mathrm{f}}^{\bar{\nu}}[\cdot] - \hat{\mathcal{P}}_\mathrm{s} \left[[\cdot]d_{\sigma_\mathrm{f}}^{\bar{\nu}} \right] \rbrace, \\
    \hat{\mathcal{B}}_j[\cdot] &=(-1)^{\delta_{p,-}}
    \lbrace \eta_{\alpha,h}^{\nu}d_{\sigma_\mathrm{f}}^\nu[\cdot] + (\eta_{\alpha,h}^{\bar{\nu}})^* \hat{\mathcal{P}}_\mathrm{s} \left[[\cdot]d_{\sigma_\mathrm{f}}^{\nu} \right] \rbrace,    
\end{aligned}
\end{equation}
in terms of the parity operator $\hat{\mathcal{P}}_\mathrm{s}$ whose action is defined as
\begin{equation}
    \hat{\mathcal{P}}_\mathrm{s}\left[ \rho_{\textbf{j}}^{(m, \pm)}(t) d_{\sigma_\mathrm{f}}^\nu \right] = \mp(-1)^m \rho_\textbf{j}^{(m,\pm)}(t) d_{\sigma_\mathrm{f}}^\nu.
\end{equation}
\begin{figure}[!htbp]
    \centering
    \includegraphics[width=1\linewidth]{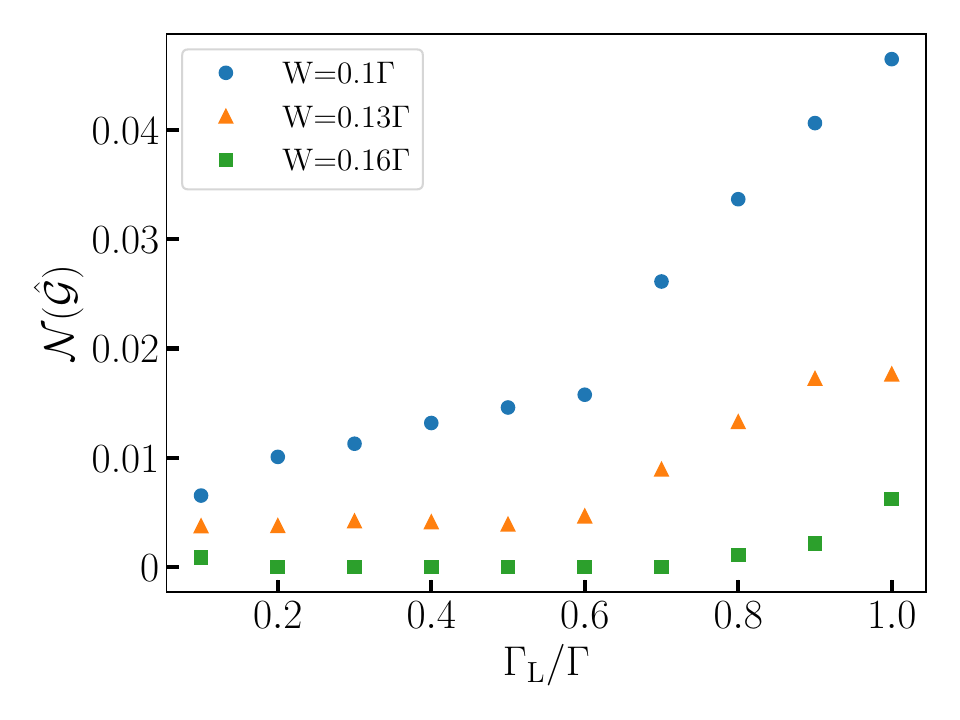}
    \caption{Non-Markovianity ($\mathcal{N}(\hat{\mathcal{G}})$) as a function of the coupling strength between the quantum dot (QD) and the left lead ($\Gamma_\mathrm{L}$) for different values of the environment's bandwidth ($W$). The truncation of the HEOM tiers and the Pad\'e series are set to $N_\mathrm{H}=4$ and $l_\mathrm{max}=2$, respectively.
    The results show that the non-Markovianity becomes less significant as $W$ increases. This trend is observed for all values of $\Gamma_\mathrm{L}$ presented. In other words, at a smaller fixed bandwidth (e.g., $W = 0.1\Gamma$), a stronger $\Gamma_\mathrm{L}$ leads to an even higher degree of $\mathcal{N}(\hat{\mathcal{G}})$.}
    \label{fig:non-Markovianity_W}
\end{figure}
In contrast, the coupling to the right lead (which will primarily serve as a detector) can be modeled using the typical assumptions valid in the weak-coupling regime.
This allows for an adequate description of the right-lead influence on the QD using the Born-Markov quantum master equation. At the same time, when strong system interactions are present, such as those arising from a large Coulomb repulsion $U$ in the QD, the conventional local Lindblad master equation can become unreliable. This unreliability can manifest in unphysical predictions, including excitations at absolute zero temperature~\cite{Beaudoin2011} or inaccurate electron occupation numbers within the QD. To overcome these problems, it is possible to use a dressed Born-Markov master equation (dBMME) which is more accurate than the conventional local Lindblad master equation~\cite{Kuo2023} to model these regimes. 
This improvement relies on restricting bath-induced transitions  among system eigenstates $\ket{\varphi_{i}}$ of $H_{\text{s}}$ having energy \blue{$\epsilon_{\mathrm{i}}$}. Technically, this corresponds to writing the Lindblad dissipator using the eigenbasis decomposition
\begin{equation}\label{eq:dress_eq}
\begin{aligned}
\partial_t\rho_{\text{s}}(t)&=-i[H_{\text{s}},\rho_{\text{s}}(t)]
+\sum_{\alpha,\sigma}
\sum_{\epsilon_{k}-\epsilon_{l}=\omega}\sum_{p=\pm}
\gamma_{\alpha,l\rightarrow k}(\omega)\times\\&
\Big\{
p\ketbra{\varphi_{k}}{\varphi_{l}}\rho_{\text{s}}^{p}(t)\ketbra{\varphi_{l}}{\varphi_{k}}
-\frac{1}{2}
\{\ketbra{\varphi_{l}}{\varphi_{l}},\rho_{\text{s}}^{p}(t)\}
\Big\},
\end{aligned}
\end{equation}
where $[\cdot,\cdot]$ and $ \{\cdot,\cdot\}$ denote the commutator and anticommutator, respectively. The transition rates in this dBMMe equation are explicitly given by
\begin{equation}
\begin{aligned}
\gamma_{\alpha,l\rightarrow k}(\omega)= 
2\pi\sum_{\nu=\pm 1}\sum_{\sigma=\uparrow,\downarrow}
\vert\langle \varphi_{k}|d^{\nu}_{\sigma}|\varphi_{l}\rangle\vert^{2}
J_{\alpha}(\omega)n^\mathrm{eq}_{\alpha}(\omega).
\end{aligned}
\end{equation}
The full equation of motion can be obtained by incorporating the strong coupling effects [described by \eq{eq:HEOM_1}] and the weak coupling effects [described by \eq{eq:dress_eq}]  to model both non-Markovian and weak interaction to the left and right leads, respectively.
This is achieved by replacing the  
$\hat{\mathcal{L}}_\mathrm{s}[\cdot]$ term in \eq{eq:HEOM_2} with the right-hand side of \eq{eq:dress_eq}.  Of course, this approximation for the right lead neglects hybridization between the system and the left lead, but we have found (compared to a full HEOM simulation by calculating the electron occupation number for both leads) that these effects are negligible.

In order to analyze the statistical properties of
the electron transport phenomena, WTD plays a crucial role. In fact, electronic currents are defined by a series of 
individual electron tunneling events occurring at random time intervals. The WTD quantifies the probability of observing an electron transfer in the detector electrode at a specific time $t + \tau$, given that an electron was previously detected in the same electrode at an earlier time $t$.
Given its experimental accessibility and also for analytical convenience, we now compute the WTD around the steady state of this model we have described.
The WTD in the steady state is derived as follows \cite{Brandes2008, Rudge2016_1, Rudge2016_2, Ptaszynski2017, Kleinherbers2021, Fu2022}
\begin{equation}
    \mathcal{W}(\J;\tau) = \frac{\tr \left[\J e^{\hat{\M}_0t}\J \rho_\mathrm{st} \right]}{\tr \left[\J\rho_\mathrm{st}\right]}.
\end{equation}
In this equation, $\rho_\mathrm{st}$ is the steady state of the auxiliary density operator (ADO) space, $\J$ is a superoperator encoding the quantum jumps in the dynamics, and $\hat{\M}_0 = \hat{\M}-\J$ is the superoperator evolving the system in the absence of quantum jumps.
Assuming conditions in which the right lead only acts as a sink for electrons, using \eq{eq:dress_eq}, the quantum jump operator reads
\begin{equation}\label{eq:jump_operator}
\begin{array}{lll}
    \J[\rho_\mathrm{s}] 
    & = &\displaystyle\sum_{\epsilon_k-\epsilon_l = \omega} \gamma_{\alpha,l\rightarrow k}(\omega)\ketbra{\varphi_{k}}{\varphi_{l}}\rho_{\text{s}}(t)\ketbra{\varphi_{l}}{\varphi_{k}}\\
    & \equiv& \displaystyle\sum_{i=1}^4 \J_{i}[\rho_\mathrm{s}]\;,
\end{array}
\end{equation}
where the definition in the last line is introduced to highlight specific electron transfer processes from the QD to the right lead.
Explicitly,
\begin{equation}\label{eq:each_jumps}
\begin{array}{lll}
   \J_{1}[\rho_\mathrm{s}] & =& \displaystyle\gamma_{\mathrm{R}, \uparrow \rightarrow 0}(-\epsilon) \ket{0} \bra{\uparrow}\rho_\mathrm{s} \ket{\uparrow} \bra{0} \\
    \J_{2}[\rho_\mathrm{s}] &= &\displaystyle \gamma_{\mathrm{R}, \downarrow \rightarrow 0}(-\epsilon) \ket{0} \bra{\downarrow}\rho_\mathrm{s} \ket{\downarrow} \bra{0} \\
   \J_{3}[\rho_\mathrm{s}]  & =&\displaystyle\gamma_{\mathrm{R}, \uparrow \downarrow \rightarrow \downarrow}(-U) \ket{\downarrow} \bra{\uparrow \downarrow}\rho_\mathrm{s} \ket{\uparrow \downarrow} \bra{\downarrow} \\
   \J_{4}[\rho_\mathrm{s}]  & =&\displaystyle\gamma_{\mathrm{R}, \uparrow \downarrow \rightarrow \uparrow}(-U) \ket{\uparrow} \bra{\uparrow \downarrow}\rho_\mathrm{s} \ket{\uparrow \downarrow} \bra{\uparrow} \;,
    \end{array}
\end{equation}
so that $\J_i$ describes transitions from the single-occupancy states $\ket{\sigma}$ ($\sigma = \uparrow, \downarrow$) to the vacuum $\ket{0}$ for $i=1,2$, and from the double-occupancy state $\ket{\uparrow \downarrow}$ to one with single-occupancy for $i=3,4$.

We can also define WTD for a specific electron transfer process, described by the quantum jump superoperator $\J_i$, with
\begin{equation}\label{eq:wtd_jp}
    w_{i}(\tau)\equiv\W(\J_i;\tau) =\frac{\tr\left[\J_i e^{\hat{\M}_0^i \tau} \J_i \rho_\mathrm{st}\right]}{\tr \left[ \J_i \rho_\mathrm{st} \right]},
\end{equation}
where $\hat{\M}_0^i = \hat{\M} - \J_i$ represents the superoperator governing the evolution without the specific jump process $\J_i$.

Since the left lead is allowed to exhibit non-Markovian effects, it is useful to introduce a measure to quantify the degree of non-Markovianity. 
\begin{figure*}[!htbp]
    \centering
    \includegraphics[width=1\linewidth]{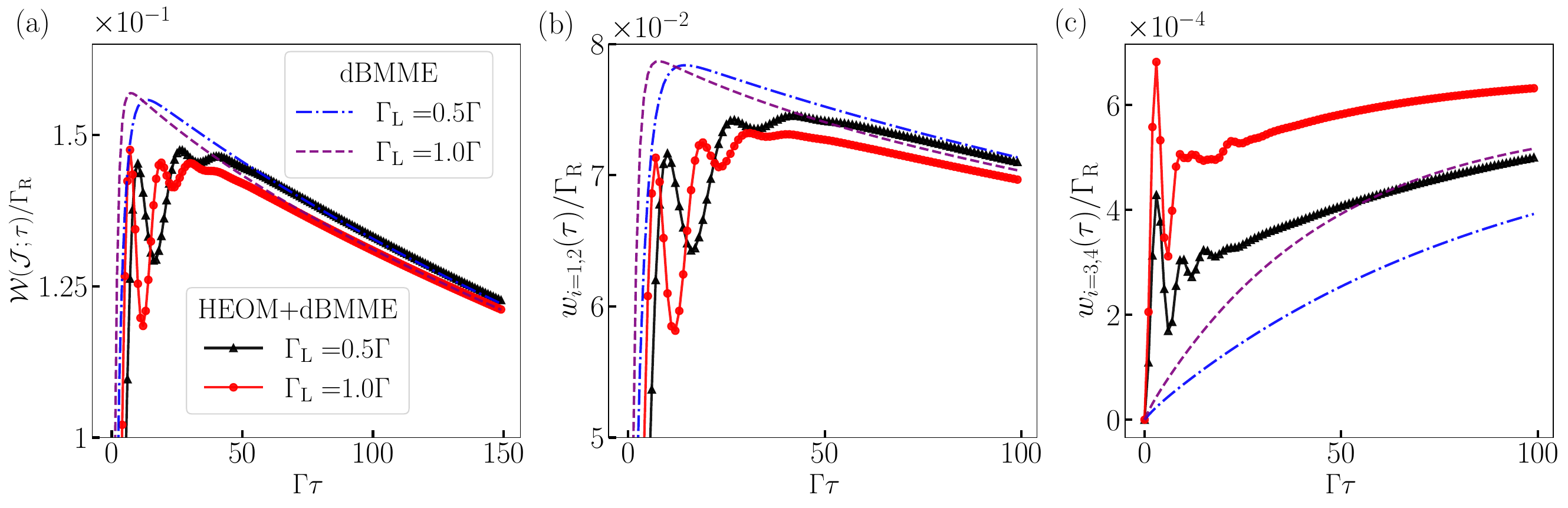}
    \caption{ (a) $\W(\mathcal{J};\tau)$ calculated by using HEOM+dBMME method (solid) and dBMME (dashed). In the strong coupling regime $(\Gamma_\mathrm{L} > ...)$, dBMME fails to capture short-time oscillations. (b), (c) Individual electron WTDs, $w_{i}(\tau)$, at $\Gamma_\mathrm{L} = 0.5 \Gamma$ and $\Gamma_\mathrm{L} = \Gamma$. HEOM+dBMME reveals significant short-time oscillations in all $w_{i}(\tau)$  – a feature absent in dBMME results. Electrons predominantly emit from single occupation states ($\ket{\uparrow}$ or $\ket{\downarrow}$, related to $\J_{1,2}$) rather than the double occupation state ($\ket{\uparrow \downarrow}$, $\J_{3,4}$)
    }
    \label{fig:different_jump}
\end{figure*}
In a Markovian process, information flows unidirectionally from the system to the environment, leading to increasingly indistinguishable quantum states over time. Conversely, non-Markovian behavior allows for information to flow back to the system, potentially overcoming this loss of distinguishability. 
The trace distance, defined as
\begin{equation}\label{eq:TD}
    D(\rho_1, \rho_2) 
    = \frac{1}{2}\mathrm{Tr}\left[ \sqrt{(\rho_1 - \rho_2)^\dagger (\rho_1 - \rho_2)} \right],
\end{equation}
quantifies the distinguishability between two quantum states, $\rho_1$ and $\rho_2$. From \eq{eq:HEOM_1}, the reduced state of the QD system is defined as
\begin{equation}
\begin{aligned}
    \rho_s(t) 
    &= \rho^{(0,+)}(t)= \hat{\mathcal{G}}(t)[\rho^{(0,+)}(0)],
\end{aligned}
\end{equation}
where $\hat{\mathcal{G}}(t)=\mathrm{exp}(\hat{\M}t)$ is the superoperator which propagates all ADOs.
For a Markovian process, the distinguishability $D(\rho_{s1}(t), \rho_{s2}(t))$ between two different initial states $s1$ and $s2$ of the QD monotonically decreases over time.
However, for non-Markovian systems,
a non-monotonic behavior emerges as a result of information backflow.
As a consequence, we can quantify the degree of non-Markovianity by considering the difference in distinguishability with respect to the Markovian case for all possible initial states as
\begin{equation}\label{eq:nonMark}
\begin{aligned}
    \mathcal{N}(\hat{\mathcal{G}}) = 
    &\max_{\rho_{s1,s2}(0)}\sum_i [D(\rho_{s1}(b_i), \rho_{s2}(b_i))\\
    &-D(\rho_{s1}(a_i), \rho_{s2}(a_i))].
\end{aligned}
\end{equation}
Here the index $i$ labels the times intervals $(a_i, b_i)$ during which the rate of change of the trace distance  is positive, indicating an increase in distinguishability due to non-Markovian information flow~\cite{Breuer2009}, i.e., 
\begin{equation}
    \frac{d}{dt}D(\rho_{s1}(t),\rho_{s2}(t))>0\;,
\end{equation}
for $t\in (a_i, b_i)$. In the following section, we use these technical tools to analyze the electronic transport properties of the model defined in Eq.~(\ref{eq:total_Ham}).

\section{Results}\label{sec:result}

We begin with an analysis of the full WTD for different values of the coupling strength to the bath $\Gamma_\mathrm{L}$ with a constant on-site Coulomb interaction energy of $U = 1\Gamma$, and for a system initially at equilibrium with chemical potential  $\mu=0$. Here and throughout the article, we use $\Gamma$ as a reference scale for all energy parameters.
In \fig{fig:WTD_coupling_spin} we show the short-time behavior of the WTD, where we further set 
$k_\mathrm{B}T=0.5\Gamma$, and constrained the energy of the QD level to $\epsilon = -0.1\Gamma$

\subsection{WTD reveals non-Markovian signature}

To compute the distinguishability of the states during the dynamics, we used Eq.~(\ref{eq:nonMark}) by maximizing over all pairs of initial conditions taken from the set $\left\{\ket{0}, \ket{\uparrow}, \ket{\downarrow}, \ket{\uparrow \downarrow}, \ket{+}, \ket{-} \right\}$ 
where $\ket{\pm}=1/\sqrt{2}\left(\ket{\uparrow} \pm \ket{\downarrow} \right)$.
As expected, non-Markovian effects become
increasingly pronounced as the system transitions into the non-Markovian regime. This observation is evident in Fig.~\ref{fig:non-Markovianity_W}, where stronger coupling strengths ($\Gamma_{\text{L}}$ ranging from $0.1\Gamma$ to $\Gamma$) lead to an increase in distinguishability.
We observe that this feature is absent when the HEOM is replaced with a Markovian master equation to model the strong coupling to the left lead,  demonstrating its inability to capture non-Markovian effects.
On the other hand, the result using the HEOM  further shows that this effect is particularly pronounced when the lead bandwidths ($W_\mathrm{L,R} = W$) are narrower, specifically for $W < 0.16\Gamma$. This suggests that narrower fermionic bandwidths can amplify non-Markovian effects.

To analyze the influence of the system-bath coupling strength on non-Markovianity, we now
investigate the waiting-time distribution (WTD), $\W(\J;\tau)$, across different
values of $\Gamma_{\text{L}}$ at constant
bandwidth 
$W_\mathrm{L,R}=W=0.1\Gamma$ for the leads.
As shown in Fig.~\ref{fig:WTD_coupling_spin}(a), a key feature emerges in the behavior of  $\W(J;\tau)$.
In fact, in a regime where $\Gamma_\mathrm{L}>0.1\Gamma$, oscillations in the waiting time distribution emerge.
Interestingly, the frequency of these oscillations increases for stronger interactions between the system and the left lead.
We remark that, here, this behavior is driven by system-bath interactions rather than the presence of coherent interactions \cite{Michele2013,Weymann2020}, or spin-state transitions within the system \cite{Sothmann2014, Krzysztof2017}.

To further describe the oscillatory behavior of the  WTD, we now consider its one-sided Fourier transform
\begin{equation}
    \W(\J;\omega)=\left| \int_0^\infty d\tau e^{i \omega \tau} \W(\J;\tau) \right|,
\end{equation}
which can be written as \cite{Kleinherbers2021}
\begin{equation}\label{eq:WTD_fourier}
    \W(\J;\omega)=\left| 
    \frac{\tr \left[ \mathcal{J}\left(i \omega \id - \hat{\mathcal{M}}_0 \right)^{-1} \mathcal{J} \rho_\mathrm{st}  \right]  }{ \tr \left[ \mathcal{J} \rho_\mathrm{st} \right] } \right|.
\end{equation}
We analyze this quantity in  Fig.~\ref{fig:WTD_coupling_spin}(b) where we observe an overall shift in the oscillation frequencies as $\Gamma_\mathrm{L}$ increases. This justifies the origin of this shift with the emergence of a stronger system-bath hybridization. We note that an equivalent behavior is also present using a simplified model in which the impurity is replaced by a single charge (see \apx{Appendix_single_charge} for details).

To confirm the role of the non-perturbative effects in capturing the WTD oscillations, in Fig.~\ref{fig:different_jump}, we compare the results when either the HEOM or the dBMME is used to describe the strong interaction to the left lead. As expected, the dBMME approach is not able to correctly model the non-perturbative effects characterizing the coupling to the left lead, thereby failing to reproduce the  WTD oscillations shown by the HEOM.

In order to refine this analysis to find the dominant processes in the electronic transport, we now  calculate the individual waiting time distributions $w_{i}(\tau)$ for each quantum jump defined in \eq{eq:jump_operator}. These are the WTD for electrons tunneling out of the QD from its single-occupancy states ($\ket{\sigma=\uparrow, \downarrow}$, for $i=1,2$) and the double-occupancy state ($\ket{\uparrow \downarrow}$, for $i=3,4$).

In Fig.~\ref{fig:different_jump}(b) and (c), we observe that the peak probability of $w_{i=1,2}(\tau)$ is significantly higher than that of $w_{i=3,4}(\tau)$. This indicates a preference for the QD to occupy single-occupancy states, likely due to the large on-site Coulomb repulsion energy ($U$) disfavoring double occupancy. In other words, tunneling to the right lead is dominated by events involving the QD single occupancy states.
However, the presence of oscillatory behavior for all $w_{i}(\tau)$ suggests that events involving single-to-empty and double-to-single state transitions are both influenced by the coherent effects characterizing the interaction to the left lead. In addition to the identical jump events in the WTD, we also investigated components of $\W(\J;\tau)$  incorporating two distinct quantum jump events. These components exhibited only minor contributions to $\W(\J;\tau)$, implying weak correlations between different jump types. The detail is presented in \apx{Appendix_dW}.
It is interesting to further analyze the relation between these non-Markovian features and the emergence of entanglement between the QD and the bath electrons which leads to the Kondo effect.

\begin{figure}[!htbp]
    \centering
    \includegraphics[width=1\linewidth]{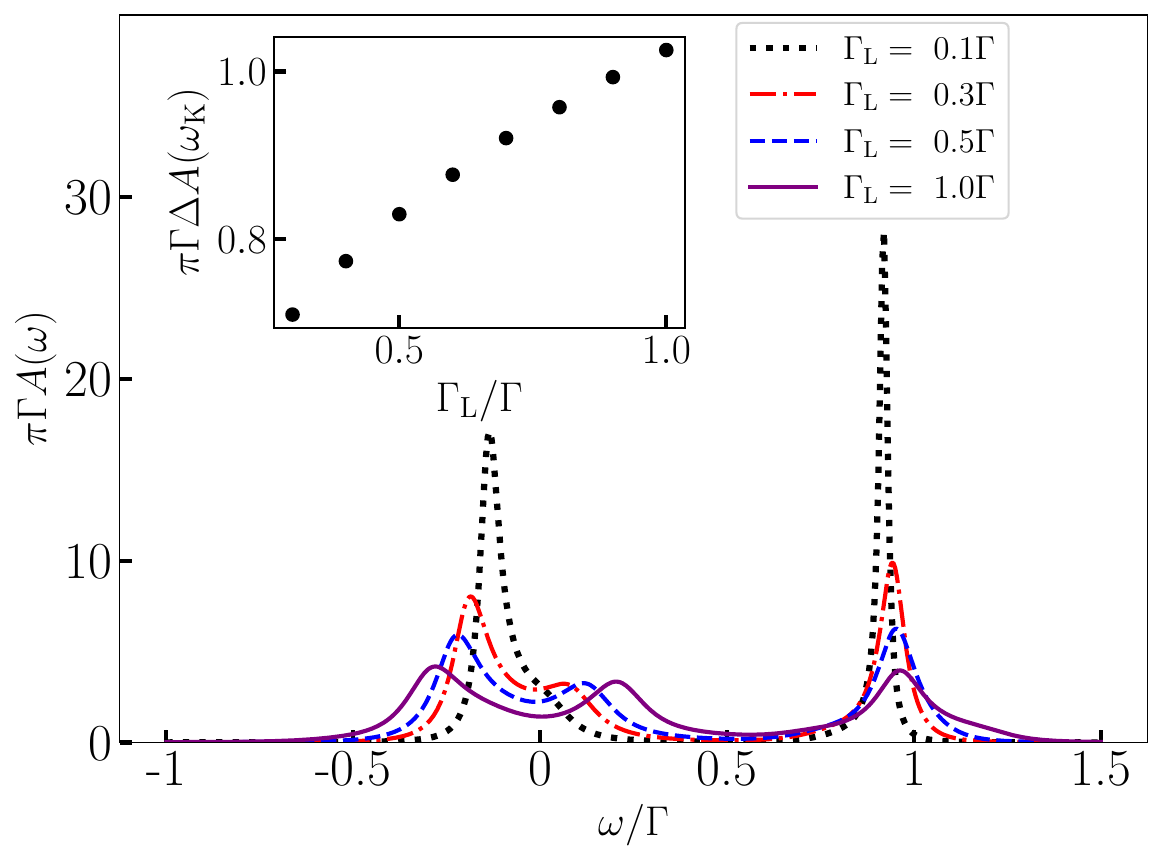}
    \caption{Density of states and Kondo resonance dependence on coupling strength. $A(\omega)$ is the Spin-up electron density of states $A(\omega)$. Two Hubbard peaks appear, along with a central Kondo peak for $\Gamma_\mathrm{L} \geq 0.3\Gamma$. The Kondo peak signifies an equilibrium many-body entangled state between the QD and left lead, and its intensity increases slightly with coupling strength (inset). Inset: $\Delta A(\omega_\mathrm{K})$, the difference in Kondo peak intensity between $k_\mathrm{B}T = 0.5 \Gamma$ and $k_\mathrm{B}T = 5\Gamma$, as a measure of Kondo resonance. Larger $\Delta A(\omega_\mathrm{K})$  correlates with stronger coupling and higher-frequency oscillations in $\W(\mathcal{J};\tau)$}
    \label{fig:DOS}
\end{figure}

\begin{figure*}[!htbp]
    \centering
    \includegraphics[width=1\linewidth]{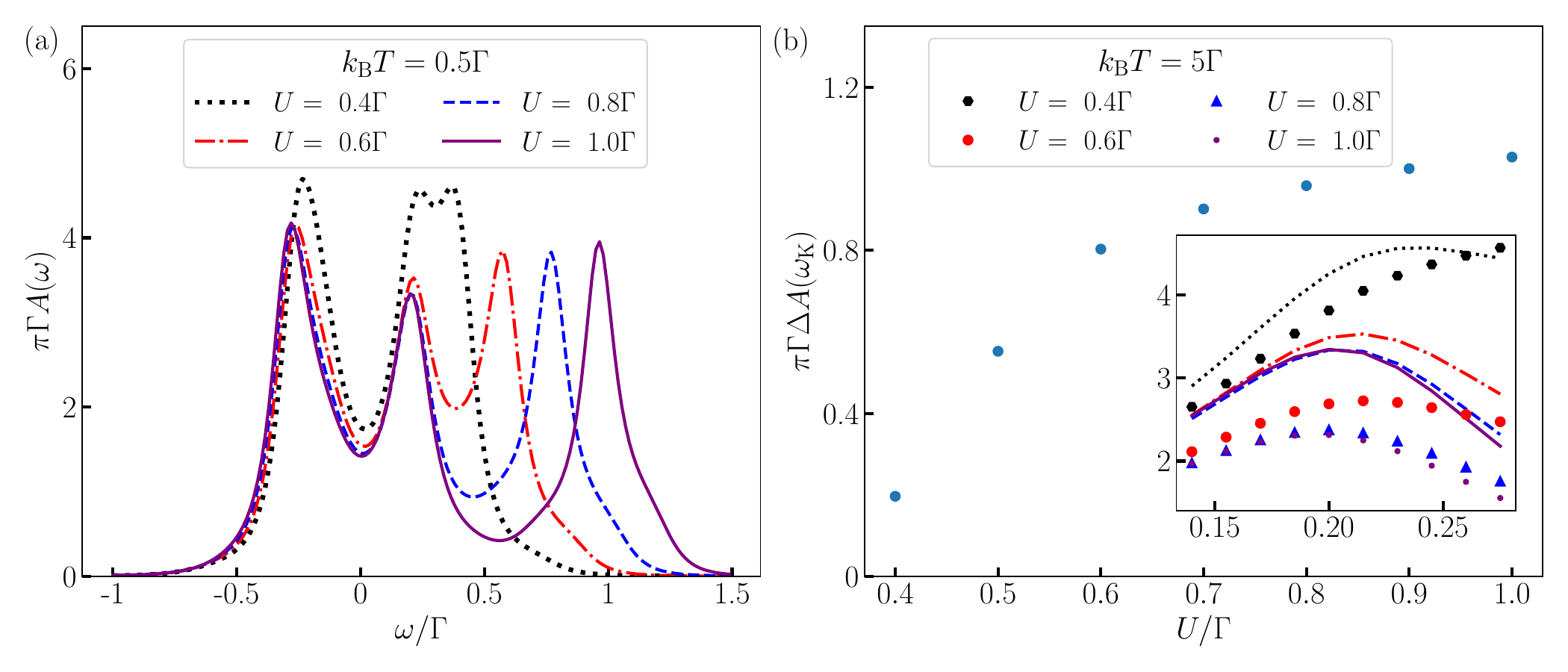}
    \caption{(a) Density of states (DOS) for varying Coulomb repulsion energies, $U$. (b) Difference in Kondo peak intensity between $k_\mathrm{B}T_1 = 0.5\Gamma$ and $k_\mathrm{B}T_2 = 5\Gamma$, a measure of Kondo resonance strength.  Larger differences correspond to stronger Kondo resonance at larger $U$. Inset: Kondo peaks at both temperatures (curves: $k_\mathrm{B}T=0.5\Gamma$; scatters: $k_\mathrm{B}T=5\Gamma$).
    }
    \label{fig:difference_DOS}
\end{figure*}

\subsection{WTD and the Kondo correlation}
To analyze the interplay between non-Markovianity and Kondo physics,
we introduce the DOS of the QD, which is an important quantity to characterize the electronic behavior of the system~\cite{Nicolas2009}.
The DOS of the QD
\begin{equation}\label{eq:DOS}
\begin{aligned}
    \pi A_\sigma (\omega)
    &=
    \mathrm{Re}\left\{ \int_0^\infty dt 
    \left\langle \{d_\sigma(t), d_\sigma^\dagger(0)\}\right\rangle e^{i\omega t} \right\}
    ,
\end{aligned}
\end{equation}
depends on system correlations which can be computed using the parity-dependent HEOM approach~\cite{Mauro2022,Huang2023} in Eq.~(\ref{eq:HEOM_1}) and  Eq.~(\ref{eq:dress_eq}).
Here, $\langle\hat{O}\rangle$ denotes the expectation value of the system operator $\hat{O}$. In the absence of an applied magnetic field, the QD exhibits spin-independent DOS. Therefore, in the following, we will focus on the DOS of the spin-up state, i.e., on the quantity $A(\omega)\equiv A_{\sigma=\uparrow}(\omega)$.
As shown in \fig{fig:DOS}, the DOS exhibits two Hubbard peaks located, for weak system-bath coupling, at the resonant energies $\epsilon$ and $\epsilon+U$ corresponding to singly and double occupied states, respectively. At strong coupling to the left lead, a central Kondo resonance peak appears as a manifestation of  the equilibrium many-body entanglement between single electron states in the system and the electrons in the reservoir continuum.

Increasing $\Gamma_{\text{L}}$ slightly enhances and blue-shifts the Kondo peak. However, this increase is subtle due to the proximity of the single-occupation distribution to the Kondo resonance.  This proximity can be attributed to the central peak height. 
Since Kondo correlations are highly temperature-sensitive (thermal fluctuations readily disrupt them~\cite{Costi2000,Kouwenhoven2001,Liang2002,Zhang2013}), increasing the lead temperature rapidly suppresses the DOS Kondo peak~\cite{Nagaoka2002,Cheng2015,Shiono2019,Kuo2023}. 

To isolate the true Kondo correlation within the DOS, we examine how the Kondo peak diminishes as the temperature exceeds the Kondo temperature (where the DOS central peak height remains fixed). We compare a higher temperature, $T_{2}=5\Gamma$, to a lower one, $T_{1}=0.5\Gamma$, and calculate the difference in Kondo peak height:

\begin{equation}
    \Delta A(\omega_\mathrm{K}) =  A(\omega_\mathrm{K},T_1) - A(\omega_\mathrm{K},T_2),
\end{equation}
where $\omega_\mathrm{K}$ is the Kondo peak frequency. As $\Gamma_\mathrm{L}$ increases, $\Delta A(\omega_\mathrm{K})$ also increases, signifying a stronger Kondo correlation induced by the enhanced QD-lead coupling (see the inset in Fig.~\ref{fig:DOS}). 

In this specific instance, one can quantify the strength of the Kondo correlation from the properties of the DOS. However, in cases when the dependence of the Kondo-peak height with respect to the lead-coupling is difficult to resolve, oscillations in the WTD  could serve as an alternative indicator to quantify the effects of system-bath non-Markovianity on the Kondo correlation.

Intriguingly, when the SIAM's repulsion energy $U$ is significantly reduced, its WTD behavior exhibits similarities to that of the coupled charge-lead system. This observation highlights the crucial role of the repulsion energy in differentiating the SIAM's behavior from the charge-lead system. Moreover, the repulsion energy $U$ contributes to the formation of a localized magnetic moment at the QD site by 
{energetically penalizing the occupation of the same site with a second electron.}
This localized moment then interacts with the conduction electrons in the lead, giving rise to the Kondo correlations.
\begin{figure*}
    \centering
    \includegraphics[width=1\linewidth]{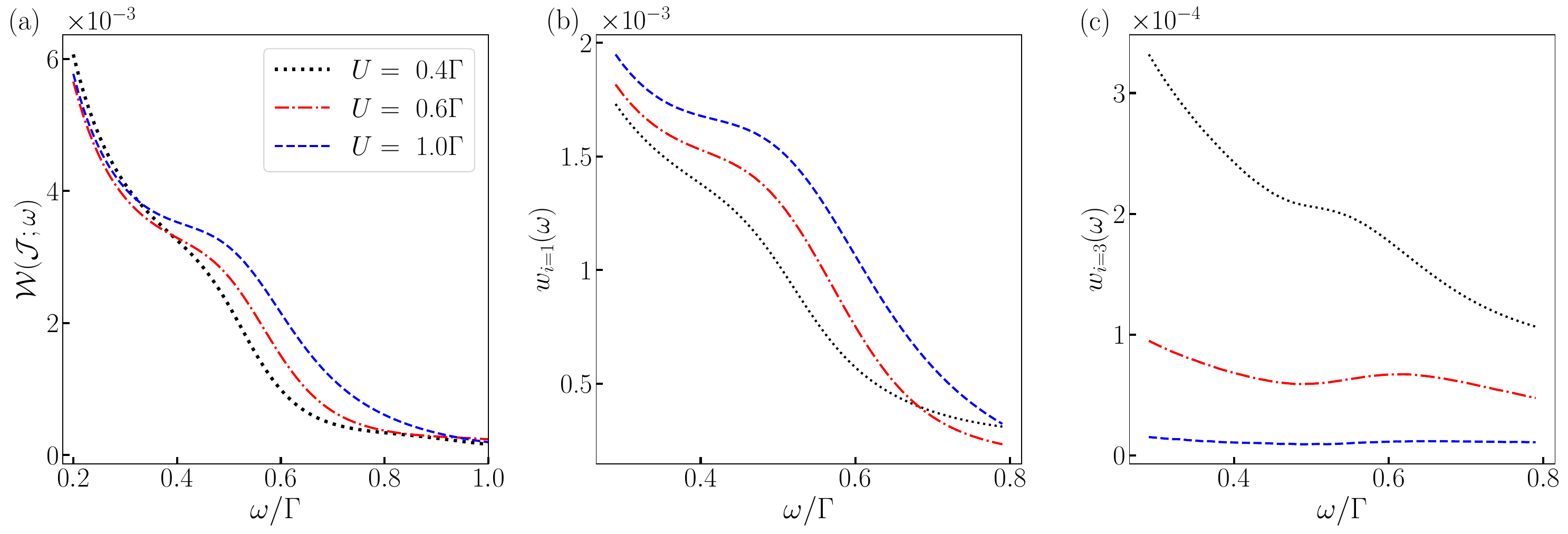}
    \caption{Fourier-transformed WTD for different $U$. (a) $\W(\J;\omega)$.  The oscillation amplitude increases with stronger Coulomb interaction $U$. (b), (c) Individual Fourier-transformed WTDs $w_{i=1}(\omega)$ and $w_{i=3}(\omega)$, showing enhanced oscillations for $w_{i=1}(\tau)$ and suppressed oscillations for $w_{i=3}(\tau)$ with increasing $U$.
    }
    \label{fig:difference_U_WTD}
\end{figure*}
As we increase $U$ from $0.4\Gamma$ to $\Gamma$ within the SIAM system, we observe not only the expected blue shift of the positive frequency Hubbard peak, but also a reduction in the central peak's height [Fig.~\ref{fig:difference_DOS}(a)]. Importantly, this decrease in the central peak is primarily due to the diminished contribution of double-occupancy distribution. The positive-frequency Hubbard peak significantly overlaps the central peak at smaller $U$ values. Consequently, it becomes challenging to quantify the Kondo peak solely through DOS analysis. However, as shown in Fig.~\ref{fig:difference_DOS}(b), $\Delta A(\omega_\mathrm{k})$ increases at  larger $U$ values. This trend indicates stronger Kondo resonances in the presence of a more intense Coulomb repulsion energy.
This is illustrated in the inset of Fig.~\ref{fig:difference_DOS}(b) which compares the DOS around the Kondo peak frequency at two temperatures, $T_1$ and $T_2$. At the higher temperature ($T_2=5\Gamma$), the Kondo peak is suppressed. Consequently, the difference $\Delta A(\omega_\mathrm{k})$ directly reflects this change in the Kondo resonance's strength.

To gain more insight, we examine the influence of $U$ on the WTD. 
By defining $w_i (\omega)$ as the Fourier-transform of $w_i(\tau)$,  we observe that an increase in $U$ causes an amplification of the oscillation width for  $w_{i=1}(\omega)$ and a suppression for $w_{i=3}(\omega)$, see \fig{fig:difference_U_WTD}(a) and \fig{fig:difference_U_WTD}(c), respectively. This is due to the fact a strong Coulomb blockade reduces the probability of double occupancy, thereby increasing the frequency of jump events from $\ket{\sigma}$ (for $\sigma = \uparrow$ or $\downarrow$) to the empty state.

Since a larger $U$ effectively increases the separation between the QD singly and doubly occupied states, it also increases  the stability of the localized magnetic moment, thereby enhancing the 
Kondo correlations by making the system more conducive to Kondo screening. 
As a consequence, this demonstrates that a larger oscillation amplitude for the quantities
$\W(\mathcal{J};\omega)$ and $w_{i=1}(\omega)$ coincide with, and can serve as an indicator of, an enhancement of the Kondo resonance.

\section{Conclusions}

In summary, we have investigated the non-Markovian transport dynamics in a single impurity Anderson model. Specifically, we considered  a  quantum dot  ultrastrongly coupled  to a lead injecting electrons and weakly coupled to a detector-lead. At short times, we observed the emergence of non-Markovianity signatures in terms of oscillations in the waiting time distribution (evaluated numerically using the hierarchical equations of motion and a Born-Markov master equation).

As the coupling to the injecting lead is increased, these oscillations become more pronounced and blue-shifted in frequency. This effect is due to the hybridization between the system and the ultra-strongly coupled lead as confirmed by the impossibility to reproduce it using a Markovian master equation.

By tuning the Coulomb repulsion in the quantum dot, we further explored the corresponding signatures in terms of the Kondo effect. In fact, by increasing the interaction between electrons in the dots, we observed the concurrent enhancement of both the Kondo resonance and the oscillations in the waiting time distribution. This sheds light on the relations and influence between non-Markovian system-bath correlations and the properties of the electron waiting time distribution and Kondo physics.
As an outlook, this work could be extended by calculating the WTD deep in the Kondo regime using fermionic pseudomodes to model strong system-bath interactions~\cite{Mauro2023}. At the same time, an analysis of the full counting statistics~\cite{Flindt2008,Stegmann2018, Brange2019,Kleinherbers2023} using the hierarchical equations of motion could further illuminate the role of non-Markovian effects in electron transport.

\section{Acknowledgement}
YNC acknowledges the support of the National Center for Theoretical Sciences and the National Science and Technology Council, Taiwan (NSTC Grants No. 112-2123-M-006-001). M.C. acknowledges support from NSFC
(Grant No. 11935012) and NSAF (Grant No. U2330401). N.~L.~is supported by the RIKEN Incentive Research Program and by MEXT KAKENHI Grant Numbers JP24H00816, JP24H00820.
\appendix
\begin{figure}[!htbp]
    \centering
    \includegraphics[width=1\linewidth]{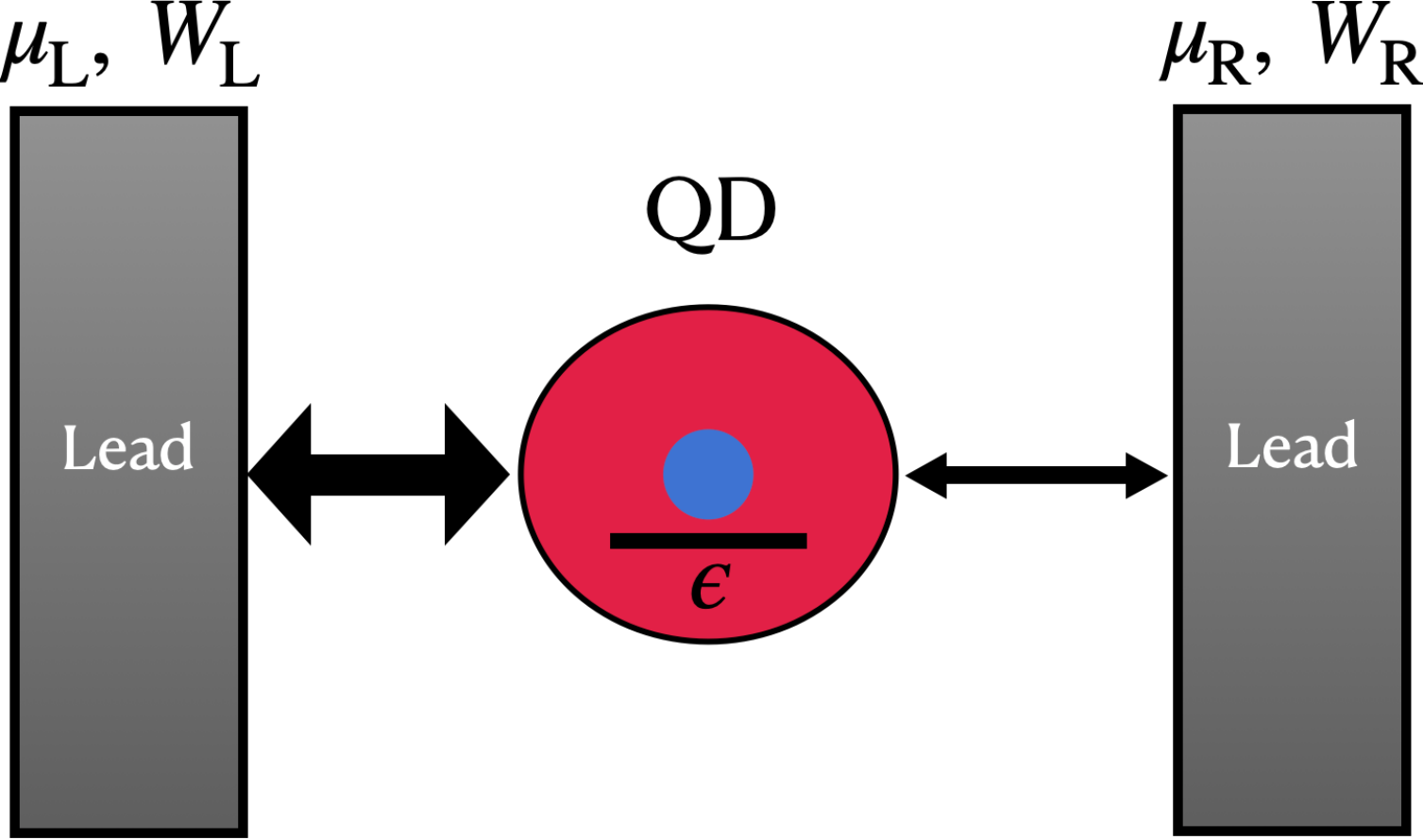}
        \caption{A quantum dot (QD) with a single energy level ($\epsilon$) is ultra-strongly coupled to its left lead (characterized by a bandwidth $W_{\text{L}}=0.1\Gamma$ and a chemical potential $\mu_{\text{L}}=0$) and weakly coupled to its right lead (characterized by  $W_{\text{R}}=0.1\Gamma$ and $\mu_{\text{R}}=0$). A detector placed in correspondence to the right lead is assumed to measure the waiting time distribution $\W(\mathcal{J};\tau)$.}
    \label{fig:SQD}
\end{figure}

\section{single charge}\label{Appendix_single_charge}

To gain more intuition over the physics analyzed in the main text, here we consider a system made out of a single charge, see \fig{fig:SQD}. By comparing the results of the HEOM and a Markovian master equation, we show that, even in this simplified setting, non-Markovian effects cause the emergence of oscillations in the WTD. We further characterize the  blue shift on the oscillation frequency using a pseudo-fermion toy-model.

\subsection{Model}

We consider a single charge coupled to two leads, as  shown in Fig.~\ref{fig:SQD} and described by the Hamiltonian

\begin{equation}
\begin{aligned}
H_\mathrm{s} &= \epsilon d^\dagger d, \\
H_\mathrm{f} &= \sum_{k,\alpha} \omega_{k,\alpha} c_{k,\alpha}^\dagger c_{k,\alpha}\\
H_\mathrm{sf} &= \sum_{k,\alpha} g_{k,\alpha}c_{k,\alpha}^\dagger d + g_{k,\alpha}^* d^\dagger c_{k,\alpha}.
\end{aligned}
\end{equation}
Here, $d^\dagger$ creates an electron in the QD, while $c_{k,\alpha}^\dagger$ creates an electron in the lead $\alpha\in\{\text{L,R}\}$. In the absence of Coulomb interactions within the QD, the quantum jump superoperator for this single charge case simplifies to $\mathcal{J}[\cdot] = d[\cdot]d^\dagger$.

\subsection{Results}
\begin{figure}[!htbp]
    \centering
    \includegraphics[width=1\linewidth]{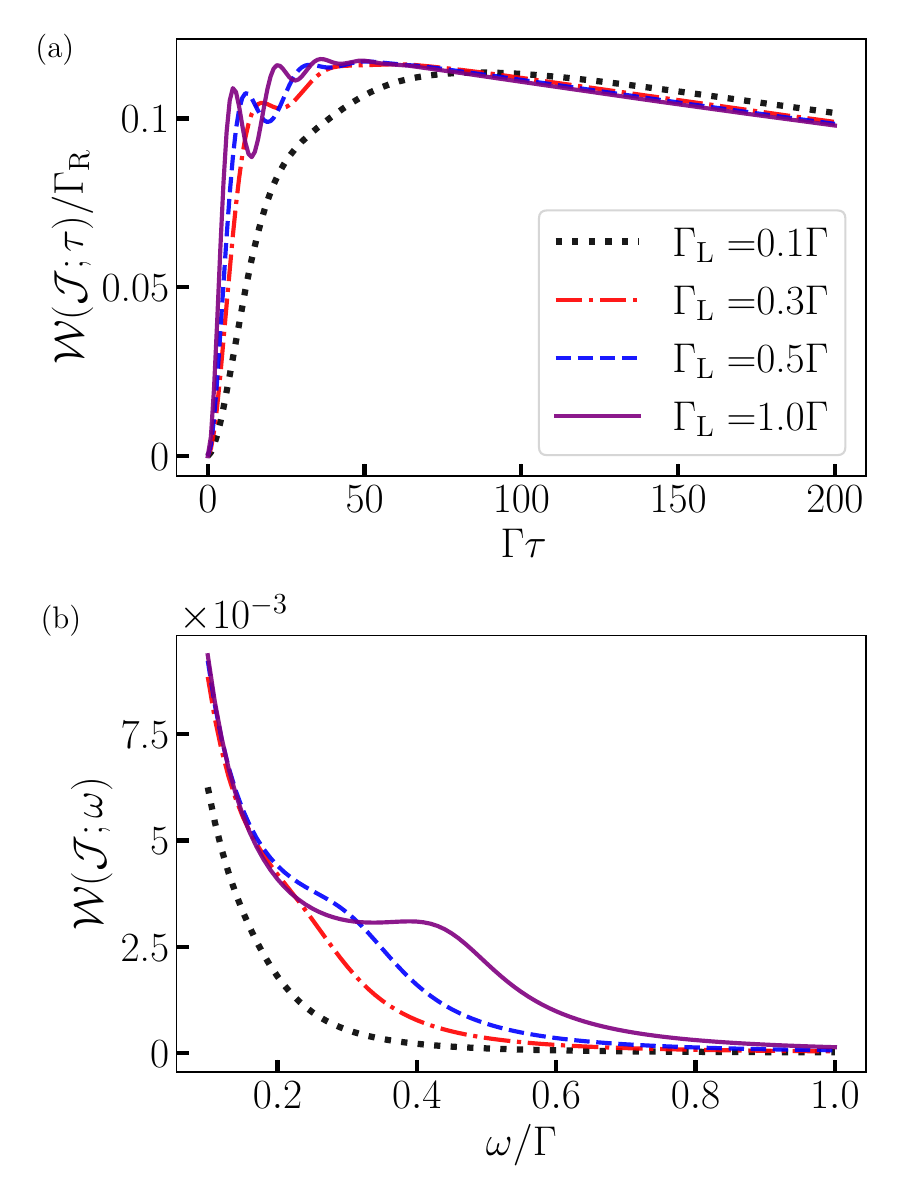}
    \caption{(a) $\W(\J;\tau)$ for different coupling strengths $\Gamma_\mathrm{L} = 0.1, 0.3, 0.5, 1 \Gamma$  to the left lead. Oscillations emerge in the short-time regime for $\Gamma_\mathrm{L} > 0.1 \Gamma$, with their frequency increasing at larger $\Gamma_\mathrm{L}$. Here, the parameters are $\epsilon = -0.1\Gamma$, $k_\mathrm{B}T = 0.5\Gamma$, and $W = 0.1\Gamma$. (b) Fourier-transform $\W(\J;\omega)$ of the WTD. Both the amplitude of the peaks and the overall blue shift are enhanced at larger $\Gamma_\mathrm{L}$, i.e. for stronger system-bath coupling.
    }
    \label{fig:single_charge_wtd}
\end{figure}
\begin{figure}[!htbp]
    \centering
    \includegraphics[width=1\linewidth]{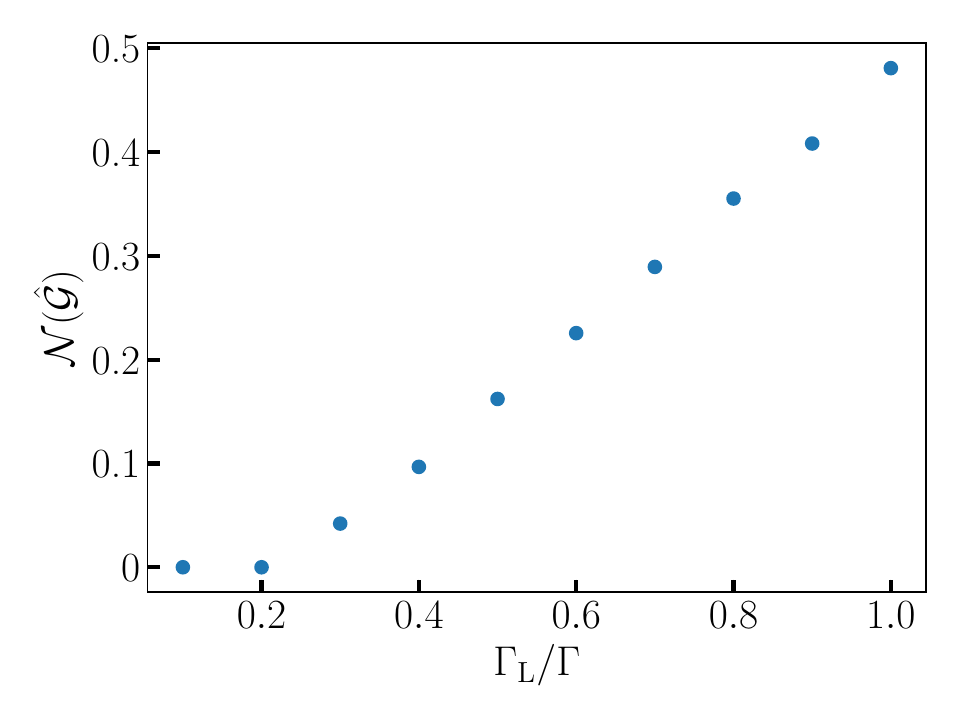}
    \caption{Non-Markovianity $\mathcal{N}(\hat{\mathcal{G}})$ as a function of the coupling strength $\Gamma_\mathrm{L}$ to the left lead . In the single charge model considered here, $\mathcal{N}(\hat{\mathcal{G}})$ increases monotonically with $\Gamma_\mathrm{L}$.}
    \label{fig:non_Mark_no_spin}
\end{figure}
In this section, we explore how the non-Markovianity of the system-bath interactions affects the waiting time distribution, $\mathcal{W}(\J;\tau)$. In particular, we will show the dependence of the oscillation amplitude and the spectral blue shift in the WTD as a function of  $\Gamma_\mathrm{L}$, in the short-time limit.

\begin{figure}
    \centering
    \includegraphics[width=1\linewidth]{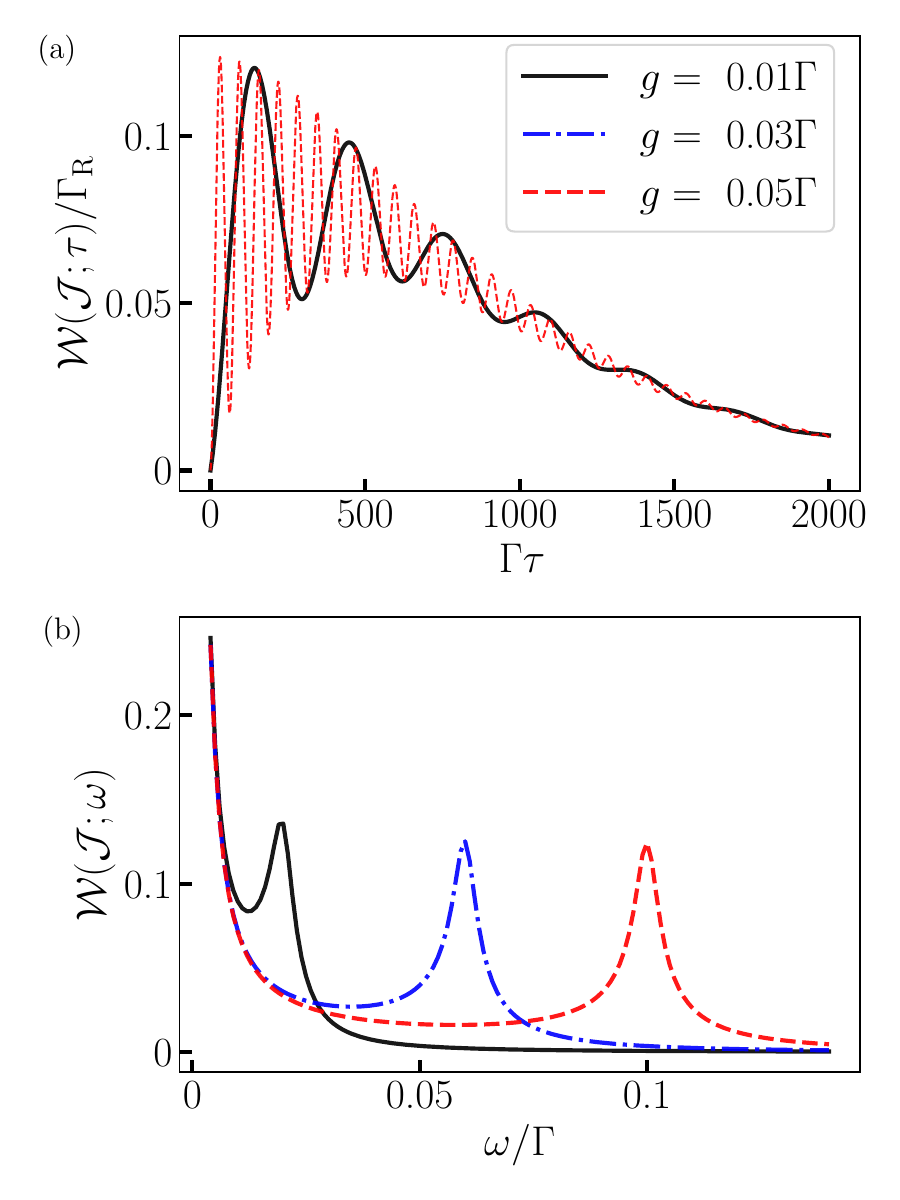}
    \caption{WTD for a pseudo-fermion toy model. In (a) we plot  $\W(\mathcal{J};\tau)$, with parameters given by $\epsilon_s = \epsilon_b = -0.1\Gamma$, $W=0.1\Gamma$, $g=0.05\Gamma$, and $k_\mathrm{B}T=0.5 \Gamma$. In (b) we plot the Fourier-transform$\W(\mathcal{J};\omega)$, as a function of the system-(pseudo-fermion) coupling strength $g$. Increasing $g$ induces a blue shift in the WTD oscillation frequency, consistently with Eq.~(\ref{toy}).}
    \label{fig:WTD_Toy_Model}
\end{figure}
As shown in \fig{fig:single_charge_wtd}(a), oscillations in $\mathcal{W}(\J;\tau)$ emerge and become increasingly pronounced at stronger values of $\Gamma_\mathrm{L}$. In Fig.~\ref{fig:single_charge_wtd}(b) we further analyze the spectral features of the WTD revealing two effects: an amplification of the oscillation amplitude and a blue shift in the oscillation frequency as $\Gamma_\mathrm{L}$ increases. 

The enhancement of the oscillation amplitude likely arises from non-Markovian effects due to stronger system-bath interactions. To support this point of view, we show that, indeed, non-Markovian effects exist and become more evident as the system-bath interaction increases which also contributes to the spectral blue-shift.
\begin{figure}[!htbp]
    \centering
    \includegraphics[width=\linewidth]{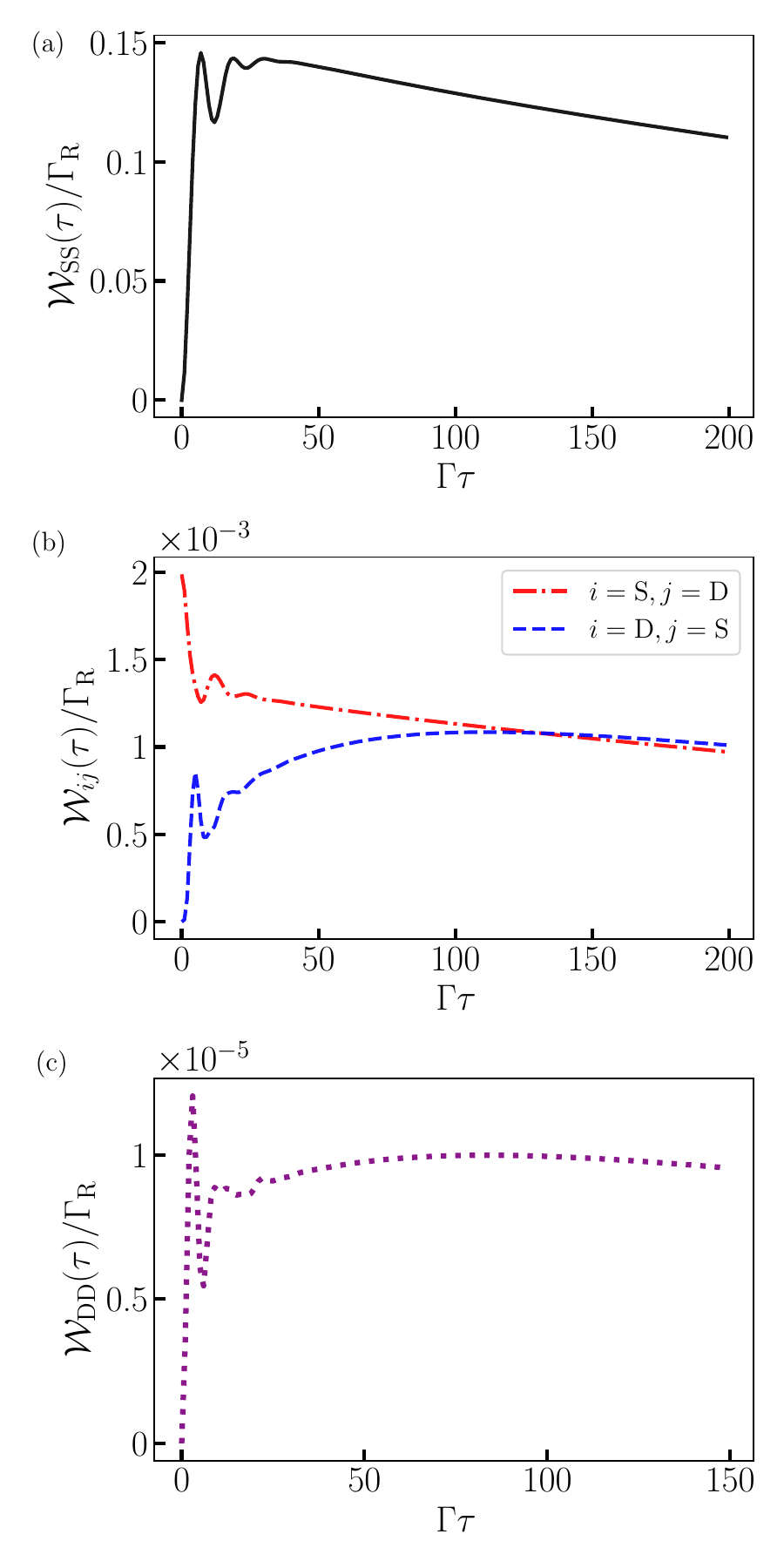}
    \caption{Contributions from
    two subsequent jumps $\J_i$ and $\J_j$ in $\W(\mathcal{J};\tau)$. 
    (a) the scale of $\W_\mathrm{SS}(\tau)$ shows that the  value of $\W(\mathcal{J};\tau)$ is dominated by contributions from events involving two $\J_\mathrm{S}$ jumps.
    (b) Here, we plot the contributions from different jump events, i.e., for $\J_i$ and $\J_j$ with $i\neq j$.
    We note that $\W_\mathrm{SD}(\tau=0)$ is non-zero since $\tr[\J_\mathrm{S}\J_\mathrm{D} \rho_\mathrm{st}]\neq 0$ allowing two electrons to jump at the same time.
    On the other hand,  $\tr[\J_\mathrm{D}\J_\mathrm{S}\rho_\mathrm{st}]=0$, which implies that $\W_\mathrm{DS}(\tau=0)=0$.
    (c) $\W_\mathrm{DD}(\tau)$. Since electrons are not likely to jump from the $\ket{\uparrow \downarrow}$ state, the contribution from  $\W_\mathrm{DD}(\tau)$ is negligible compared to other double-jump events.
    Here, $W=0.1\Gamma$, $\epsilon=-0.1\Gamma$, $U=1\Gamma$, $\Gamma_\mathrm{L}=1\Gamma$, and $\Gamma_\mathrm{R}=0.01\Gamma$.
    }
    \label{fig:Wij}
\end{figure}
We do this in \fig{fig:non_Mark_no_spin} where we plot the non-Markovianity $\mathcal{N}(\hat{\mathcal{G}})$ (which quantifies the back-flow of information from the environment to the system) as a function of $\Gamma_\text{L}$. The monotonic increase in $\mathcal{N}(\hat{\mathcal{G}})$ with $\Gamma_\text{L}$ is consistent with the enhanced oscillation amplitude in the WTD, as they both originate from the presence of stronger system-bath interaction.

We are now going to present a toy model to justify these effects more in detail.

In fact, for sufficiently strong couplings, coherent effects  can start to manifest in the energy exchange between the system and the bath. To model this, we are going  to approximate the left lead with a pseudo-fermion  coherently coupled to the system as described by the Hamiltonian

\begin{equation}
    H_\mathrm{T}^\mathrm{t} = H_\mathrm{s}^\mathrm{t} + H_\mathrm{f}^\mathrm{t} + H_\mathrm{sf}^\mathrm{t}.
\end{equation}
Here, $H_\mathrm{s}^\mathrm{t}$ characterizes the system quantum dot, $H_\mathrm{f}^\mathrm{t}$ describes both the pseudo-fermion (modeling the left lead) and the right lead (acting as a detector), and $H_\mathrm{sf}^\mathrm{t}$ is the interaction Hamiltonian between the system and both leads.

These Hamiltonians are specifically defined as follows

\begin{equation}
\begin{aligned}
H^\mathrm{t}_\mathrm{s} &= \epsilon_\mathrm{s} d_\mathrm{s}^\dagger d_\mathrm{s}, \\
H^\mathrm{t}_\mathrm{f} &= \epsilon_\mathrm{b} d_\mathrm{b}^\dagger d_\mathrm{b} + \sum_k \omega_k c_k^\dagger c_k, \\
H^\mathrm{t}_\mathrm{sf} &= g (d_\mathrm{s}^\dagger d_\mathrm{b} + d_\mathrm{b}^\dagger d_\mathrm{s})  + \sum_k g_k ( c_k^\dagger d_\mathrm{s} + d_\mathrm{s}^\dagger c_k ),
\end{aligned}
\end{equation}
in terms of the fermionic creation operators for the system $d_{i=\mathrm{s}}^\dagger$, for the pseudo-fermion $d_{i=\mathrm{b}}^\dagger$, and for the electrons in the right-lead $c_k^\dagger$. Here, $g$ is the interaction strength between the system and the pseudo-fermion.

Interestingly, this model allows to analytically estimate the frequency of the oscillations in the WTD shown in Fig.~\ref{fig:WTD_Toy_Model}(a) as

\begin{equation}
\frac{1}{2} \sqrt{(\gamma_\mathrm{e}- \gamma_\mathrm{a})^2-16g^2}, \label{toy}
\end{equation}
where $\gamma_\mathrm{e} = J_\mathrm{R}(\epsilon)(1-n^\mathrm{eq}_\mathrm{\alpha}(\epsilon))$ and $\gamma_\mathrm{a}=J_\mathrm{R}(\epsilon)n^\mathrm{eq}_\alpha(\epsilon)$. 
In fact, Eq.~\ref{toy} clearly shows the blue shift in frequency as $g$ is increased as observed in Fig.~\ref{fig:WTD_Toy_Model}(b). These findings align with the data provided in  Fig.~\ref{fig:single_charge_wtd} to provide further evidence that strong system-bath interactions lead to the emergence of both the non-Markovian oscillations in $\mathcal{W}(\J;\tau)$ and the corresponding blue shift in the frequency domain.

\section{Decomposition of $\W(\mathcal{J};\tau)$}\label{Appendix_dW}

Here, we analyze a decomposition of the  $\W(\J;\tau)$ in terms of  different jump processes for the SIAM case. In fact, the full interaction operator $\J$ encodes the effects of different processes such as the jumps
$\J_1$ and $\J_2$ of electrons from the single occupation state and the jumps $\J_3$ and $\J_4$ of electrons from the double occupation state into a single occupied one. More specifically, we can define  $\J=\J_\mathrm{S} + \J_\mathrm{D}$ in terms of $\J_\mathrm{S} = \J_1 + \J_2$ and $\J_\mathrm{D} = \J_3 + \J_4$ and decompose the WTD as

\begin{equation}
\begin{aligned}
    \W(\mathcal{J};\tau) 
    &= \frac{\tr \left[ (\J_\mathrm{S} + \J_\mathrm{D}) e^{\hat{\mathcal{M}}_0 \tau} (\J_\mathrm{S}+\J_\mathrm{D})\rho_\mathrm{st} \right]}{\tr \left[ \J \rho_\mathrm{st} \right]} \\
    &= \W_\mathrm{SS}(\tau)+\W_\mathrm{DD}(\tau)+\W_\mathrm{SD}+\W_\mathrm{DS}(\tau),
\end{aligned}
\end{equation}
where 
\begin{equation}
    \W_{ij}(\tau) =\frac{\tr\left[\J_i e^{\hat{\mathcal{M}}_0 \tau} \J_j \rho_\mathrm{st} \right]}{\tr\left[\J \rho_\mathrm{st} \right]}.
\end{equation}
In other words, $\W_{ij}(\tau)$ characterizes the effects of of the jump  $\J_j$ followed by the jump $\J_j$ after a time $\tau$.

We can now investigate $\W_{ij}(\tau)$ in the case when the impurity is at equilibrium ($\mu=0$) in the ultra-strong coupling regime to the left lead ($\Gamma_\mathrm{L}=1\Gamma$) and weak coupling to the right lead ($\Gamma_\mathrm{R}=0.01\Gamma$).
We also consider a large repulsion $U=1\Gamma$ which implies $\tr[\J_\mathrm{S}\rho_\mathrm{st}] \gg \tr[\J_\mathrm{D}\rho_\mathrm{st}]$ and justifies the relative scales  $\W_\mathrm{SS}(\tau) \gg \W_\mathrm{SD}(\tau), \W_\mathrm{DS}(\tau), \W_\mathrm{DD}(\tau)$ in  \fig{fig:Wij}. In fact, in this regime, processes for which an electron leaves the system are mostly determined by $\J_\mathrm{S}$, which leads to $\W_\mathrm{SS}(\tau) \approx \W(\mathcal{J};\tau)$ in \fig{fig:Wij}(a).
At the same time,  the fact that the jumps $\J_\mathrm{D}$ bring the system to the singly-occupied state further implies that $\tr[\J_\mathrm{S} \J_\mathrm{D}\rho_\mathrm{st}] \neq 0$, which is manifested in the non-zero value for  $\W_\mathrm{DS}(\tau=0)$ in \fig{fig:Wij}(b). Similarly,  $\J_\mathrm{S}$ characterizes jumps into the  empty state, resulting in $\tr[\J_\mathrm{D}\J_\mathrm{S}\rho_{st}]$ and $\W_\mathrm{DS}(\tau=0)=0$. 
Furthermore, \fig{fig:Wij}(c) shows  that processes which involve two consecutive jumps in $\J_\mathrm{D}$ are less likely to happen, minimizing the influence of $\W_\mathrm{DD}(\tau)$ on  $\W(\mathcal{J};\tau)$.

%


\end{document}